# Adaptable platform for trapped cold electrons, hydrogen and lithium anions and cations

L. O. A. Azevedo[1], R. J. S. Costa[1], W. Wolff[1], A. N. Oliveira[2], R.L. Sacramento[1], D.M. Silveira[1], C. L. Cesar[1*]

[1]Instituto de Física, Universidade Federal do Rio de Janeiro, Rio de Janeiro, RJ 21941–909, Brazil
[2]INMETRO, Av. Nossa Senhora das Graças, 50, Duque de Caxias, RJ 25250-020, Brazil

**Abstract**

**Cold cations, electrons and anions are ubiquitous in space, participate in star formation chemistry and are relevant to studies on the origin of molecular biology homochirality. We report on a system to generate and trap these species in the laboratory. Laser ablation of a solid target (LiH) facing a sublimating Ne matrix generates cold electrons, anions, and cations. Axial energy distributions (of $e^-$, $H^\pm$ and $Li^\pm$) peaked at 0–25 meV are obtained in a Penning trap at 90 mT and 0.5 eV barrier. Anions can be guided and neutralized with low recoil energy by near–threshold photodetachment. An immediate prospect for this $H^-$ source is to load hydrogen atoms into the ALPHA antihydrogen trap at CERN towards direct spectroscopic comparison of both conjugated species beyond 13 significant figures. The production is potentially scalable and adaptable to different species including deuterium and tritium, relevant for neutrino mass and fusion research.**

## Introduction

Sources and traps of cold negative and positive species are needed to study the low-temperature species themselves and the reactions between trapped charged particles and neutral species[1]. Crucial requirements on both sources and traps represent a significant challenge due to specific limitations and conditions on a variety of setups. Here, we report source and trap with innovative methods with an integrated mass discriminator, describing the elements and quantitatively assessing their performance. Many of the innovations and findings are of general interest, as briefly presented for diverse applications.

Measurements of the 1S–2S transition frequency in antihydrogen ($\bar{H}$) by the ALPHA collaboration at 12 significant figures[2,3] entered uncharted territory in the comparison of matter and antimatter, a test of the Charge–Parity–Time (CPT) symmetry. We foresee the need to perform laser spectroscopy of H in the same trap[4] as $\bar{H}$ to achieve aimed precisions of 15 significant figures in search of explanations for the matter–antimatter asymmetry in the Universe. In the same trap and reference frame, both species could be studied under the same conditions enabling better control over systematic effects, such as trap magnetic fields and laser power causing AC Stark shift. The $\bar{H}$ research program also involves probing the gravitational acceleration[5,6,7,8]. Antihydrogen is produced – from its constituents, antiproton and positron – and studied in challenging conditions, such as an ultra–high–vacuum (UHV) environment to avoid annihilation to background gas over many hours of stacking and laser cooling[3]. On the other hand, H has been trapped and subjected to high–precision laser spectroscopy at MIT[9] and Amsterdam[10] in traps that required a superfluid liquid helium covered cell, a condition not reachable with the $\overline{H}$ trap at 4.2 K. Therefore, innovative techniques are required for loading H in the $\overline{H}$ trap and the developments presented here can be readily adapted as a solution.

The samples of $H^-$ produced by our system are within the temperature of the antiproton and positron samples used by the ALPHA experiment for $\overline{H}$ synthesis. The $H^-$ can be produced and trapped adjacent to the main apparatus, and after the UHV condition is regained, the anions will be guided into the combined Penning and magnetic ALPHA trap. They can be further cooled by evaporative cooling by themselves or with pre–cooled electrons. Then, a laser pulse, with photon energy near the photodetachment threshold[11] of $H^-$ (0.754 eV), will neutralize the anions imparting low recoil energy. For example, a laser at 1575 nm will leave 0.2 K of recoil energy – less than the typical temperature or energy dispersion of the ion sample – to the neutral H. The fraction of resulting atoms with energy below 0.5 K will remain trapped in the superposed magnetic trap and could be detected using the sensitive technique proposed in Ref.[4].

Moreover, the developments with $\overline{H}$ research have spurred a renewed interest in hydrogen trapping and spectroscopy by many groups. New techniques to produce cold hydrogen[12-15] are being investigated with interests ranging from gravitational quantum states over a surface[12,16], to scientific metrology – in tests of Quantum Electrodynamics and proton radius puzzle[17], and search for variation of fundamental constants[18] – and to produce larger Bose–Einstein condensates[15]. The system presented offers an alternative, study model, or a proof–of–principle for some of these studies. For example, the generation of cold H from $H^-$ is the matter counterpart process from one proposed experiment[7,19] for measuring gravity with $\overline{H}$.

The present demonstration, starting from laser ablation of LiH and generating $H^-$, $Li^-$ and $Li_n H_m^-$ does not show, a priori, a specificity that would prevent it from being applicable to other simple species, such as $D^-$ and $T^-$. An example of such a specificity is the case of the MIT and Amsterdam H traps that could not trap[20] D because of its slightly higher binding energy to liquid helium. A proposal for neutrino mass measurement[21,22] requires quasi–trapped tritium for the experiment to be performed in a magnetic field. Many low–energy atomic processes involving deuterium are relevant in fusion research[23] for a copious production of $D^-$. The source described here, adapted for $T^-$ and $D^-$, might find use for these studies.

Lastly, the interaction of low energy ions and electrons with neutral atoms and molecules is important from astrophysics to the origin of biological molecules[24-26]. It would thus be desirable to have a platform to generate various cold species in a direct way. Producing and trapping slow anions is not trivial. Stray or contact electric fields can easily generate thousands of kelvins, since an energy of only 26 meV is equivalent to 300 K in temperature scale. Difference of work functions in metals can reach 1 V. Patch or domain potentials within an electrode[27] can reach 250 mV. Anions are typically created at energies[28] of many eV or keV. Various methods have been employed for further cooling ionized species: – entrainment in buffer gas[29,30]; – resistive cooling[31]; – direct sympathetic cooling[32] and LC mediated sympathetic cooling[33]; – direct laser cooling for species, such as cations[34] and proposed for molecular anions[35], with nearly closed optical transitions; – laser induced evaporative cooling of anions[36].

At the ALPHA experiment, where the $\bar{\mathrm{H}}$ trapping rate depends strongly on a low positron cloud temperature, the final cooling to ~20 K relies on evaporative cooling[37]. In summary, there are many different techniques being explored and developed to produce cold charged samples due to their scientific relevance. Here, we report on a platform to directly generate: (i) electrons with a perspective for a cold polarized electron beam; (ii) simple molecular ions at temperatures compatible with interstellar media chemistry; (iii) hydrogen anions suitable for antihydrogen research in a method adaptable to deuterium and tritium research. It may also serve as an attractive initial step towards samples in the cold and ultracold regime.

**Results**

The experimental method is based on the Matrix Isolation Sublimation (MISu) technique[38–41] – described in Methods – and variations. Similarly to the formation of molecules in the matrix[41] by the sequential implantation of different atoms, the implantation of atoms and molecules followed by electrons should result in the formation of anionic species in the matrix. As a faster variation, we found that the laser ablated species impinging onto the Ne matrix already generate low–energy electrons and atomic or molecular ions. The experimental setup and basic procedure used in this study is presented in Fig. 1a,b and described in Methods.

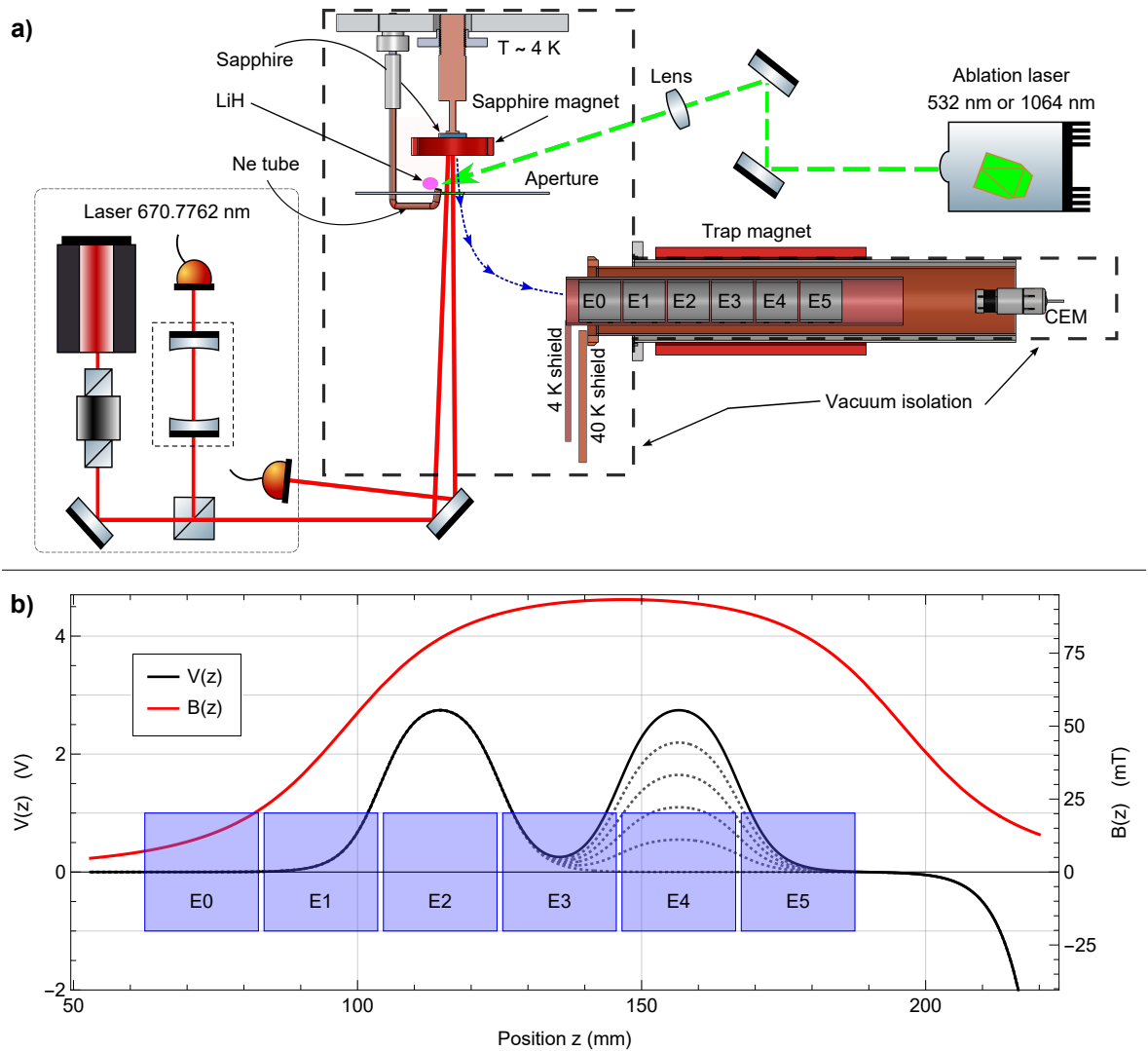

Fig. 1. Experimental setup and Penning trap field and potentials. a) A diagram of the basic setup is shown. The sapphire where the Ne matrix is deposited, as the gas is delivered through the tube, and the inner chamber are thermally anchored at the 2nd stage 4 K cryohead surrounded by a 40 K black–body shield. The spectroscopy laser monitors the matrix thickness and the Li atoms absorption. A pulsed laser at 532 or 1064 nm (in dashed green) promotes ablation from a LiH pellet imparting atoms, molecules, electrons, and ions onto the Ne matrix. Two magnets, placed around the sapphire and the trap region, are used for guiding (dotted blue arrows) and trapping the particles. An aperture avoids excess Ne gas towards the cell bottom and trap region. The Penning–Malmberg trap uses six ring electrodes (E0–E5), glued externally to a grounded copper tube that reaches 6.5 K, and is followed by a channel electron multiplier (CEM) detector. The CEM is loosely thermally anchored to the 40 K stage. b) Axial profiles of the magnetic field (red line) and a configuration, for cations, of the potentials for a dump at E4 (black solid line going in time to the lowered dotted curves) for a trap initially centered around E3.

Initial experiments involved using a custom-built time–of–flight mass spectrometer (ToF–MS) to detect the different species emanating from the sapphire. The details of this ToF–MS can be found in ref[41]. The region where the ions are extracted is large and the ToF–MS resolution is poor, but sufficient to discriminate electrons from H⁻ and Li⁻. A typical mass spectrum of the raw data and histogram is shown in Fig. 2. The extraction accelerating potential was switched on 50 µs after ablation. The electrons are detected immediately after the extraction pulse, followed by the heavier species with clear presence of H⁻, Li⁻ and molecular anions. As expected from the ablation of LiH we generated Li, H, and $Li_nH_m$ ionized species, with both signs of charge, as well as neutral atoms and molecules.

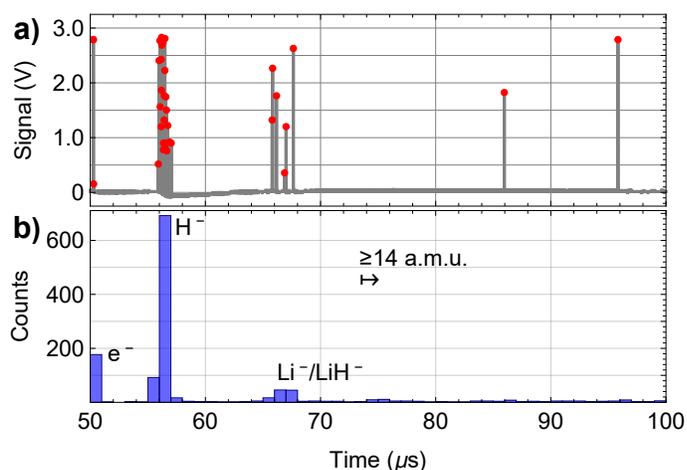

Fig. 2. Time-of-flight mass spectrometry (ToF-MS) of ions reflected from the Ne matrix after an ablation pulse. The accelerating electric field is switched on at t=50 µs after the ablation laser pulse, applied after the matrix sublimation has been initiated. The electrons appear immediately, followed by the heavier species. Data collected from the oscilloscope in a single realization is shown in (a) where the red dots are peaks identified by the LabView routine. A histogram of an accumulation of 35 runs is shown in (b) with mass identification up to Li⁻ or LiH⁻ and heavier masses ($m \geq 14$ atomic mass units due to molecules such as $Li_nH_m^-$, with n>2) appearing after t=74 µs as identified by the arrow. The fraction of counts for different species depends on timing.

In a typical MISu procedure the trapping experiment is as follows: (i) the Ne matrix is grown and laser ablation generates and implants atoms, molecules, ions and electrons into the matrix; (ii) both magnets are energized and the Ne matrix is sublimated with a heat pulse starting at time t=0, sometimes the last electrode E5 is energized from the beginning; (iii) at time $t_1$ the entrance trap electrode (E2, for example), or both E2 and E5, is switched on and capture the trapped ions; (iv) the trapping potentials are manipulated and then lowered at different times and rates and the ions detected in the CEM.

The raw data for cations of a trapping sequence is shown in Fig. 3. During the sublimation (heat pulse starts at t=0), the spectroscopy laser (purple trace) continuously scanning (at 2 kHz) around the D2 resonance in Li (670.776 nm) records the atoms' absorption after they are released from the matrix. With E5 = 5 V during sublimation, the energy barrier prevents any cation from reaching the CEM. At t=1080 μs the entrance electrode E2 is switched, during 20 μs, to 5 V closing the trap. At time t=3500 μs the exit electrode E5 is linearly brought, during 800 μs, to 0 V and the ions are detected in the CEM as shown in the blue histogram.

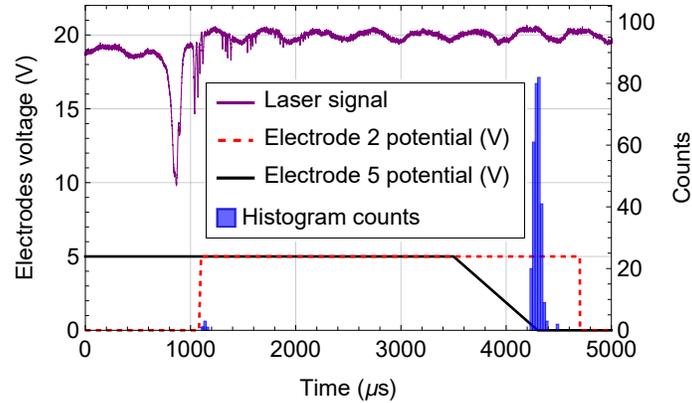

Fig. 3. Trapping of cations using the MISu procedure. A matrix is grown while 6 ablation pulses deposit atoms and charged particles into it. At t=0, a current is applied to the sapphire's NiCr film resistor sublimating the matrix along ~1100 μs. The laser transmission is shown in purple and presents information on the matrix thickness by the interference fringe and the atomic absorption (seen from 800 – 2600 μs). The E5 electrode (black trace) is set at 5 V since t=0. The E2 electrode (red dashed trace) is switched to 5 V at t=1080 μs trapping particles between E2 and E5. At t=3500 μs, E5 is linearly brought to 0 V and cations are allowed to scape towards the CEM. The cations count is shown in the blue histogram. The appearance of signal at the end of, and after, the ramp is compatible with very low energy ions, despite a small propagation delay of the ions to the detector. The small signal of ions at t~1100 μs results from the energy gained by the switching of E2 that pushes ions over the E5 barrier.

The results with a slow (2.5 ms) dump for cations shown in Fig. 4 employed the MISu procedure and one extra ablation pulse during the sublimation. The trap barrier (electrode E4) was lowered at a much lower rate than in Fig. 3, taking 2500 μs to go from 1 V to 0 V. In this case, the propagation delay to the CEM is of smaller relevance during the ramp down. An energy distribution, incorporating the counts after the ramp down into the first bin, is shown as an inset of the figure. The trap depth is not a linear function of time near the end of the ramp (notice the change in the bottom potential in Fig. 1b). The peak of the distribution is in the first bin (0 to 25 meV) which accounts for 32% of the total counts shown. Among the various manipulations we included compressing the sample under a single electrode to rid the trap of species with masses higher than 10 atomic units.

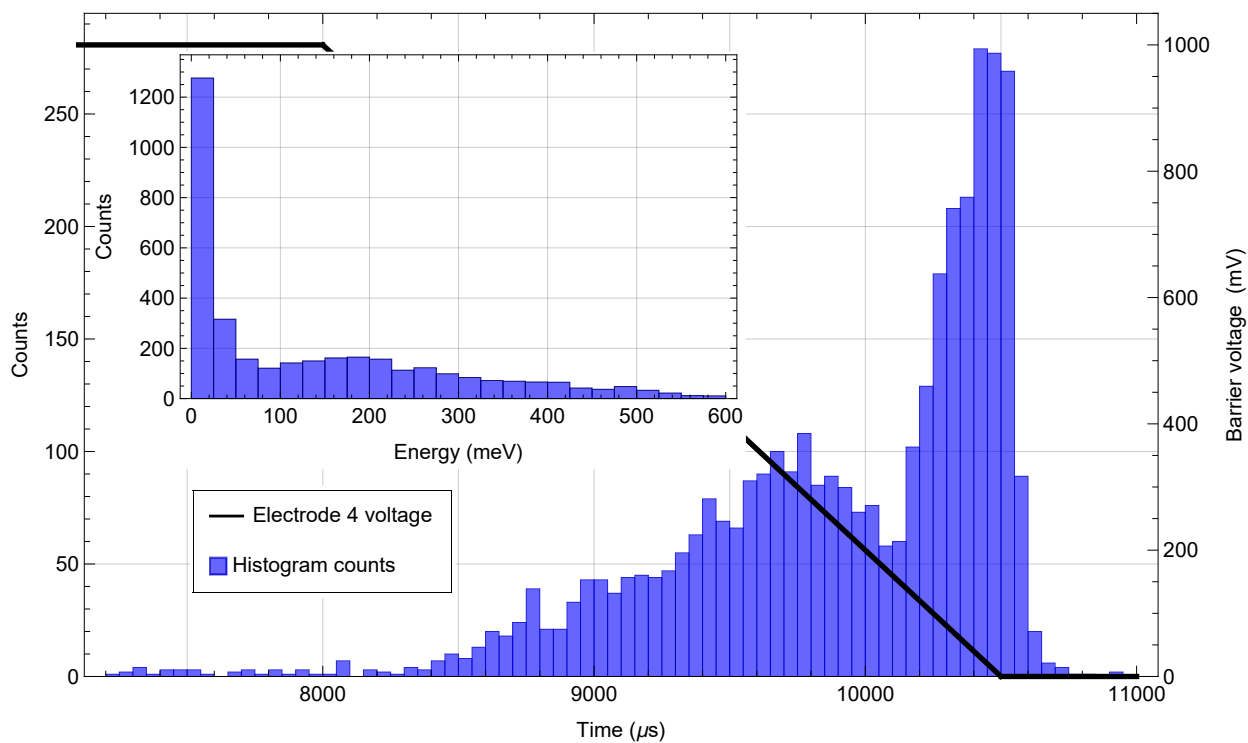

Fig. 4. Histogram of cations (H⁺ and Li⁺) released from the trap. A matrix (~1 μm) was grown over 20 s with 40 ablation pulses. The sublimation heat pulse started at t=0 resulting in absorption signal of Li atoms from t~4.5 – 8 ms. An extra ablation pulse is given at t=6140 μs and E2 rose to 5 V from 6200 – 6250 μs closing the trap between E2–E5. The potential well was later compressed, confining particles between E2–E4 and E4 was brought to 1 V. The histogram (in blue) from the final slow linear dump (from 8000–10500 μs) of E4 from 1 V is shown above in solid black. The inset shows the time distribution mapped into an energy distribution with the counts after the ramp down being incorporated in the first bin (0 – 25 meV). The data represent an accumulation of 5 experimental runs.

The MISu procedure for negative charges resulted in electrons dominating the counts. To improve the H⁻ yield, we developed a faster variation on MISu which consists of applying a single ablation pulse during the sublimation of a pure Ne matrix, and then trapping the particles. Since many laser pulses can be applied during the matrix sublimation, this variation should allow future stacking of trapped ions using a single matrix growth cycle. The results of an experiment under this variation for negative particles is shown in Fig. 5. In this case the trap was manipulated as in Fig. 4 above, going from an initial trap in E2–E5 to a compressed well in E2–E4 and then E4 slowly dumped from -2 V while E2 was kept at -5 V.

The time distribution can be folded into an energy distribution shown in the inset with bins of 25 meV.

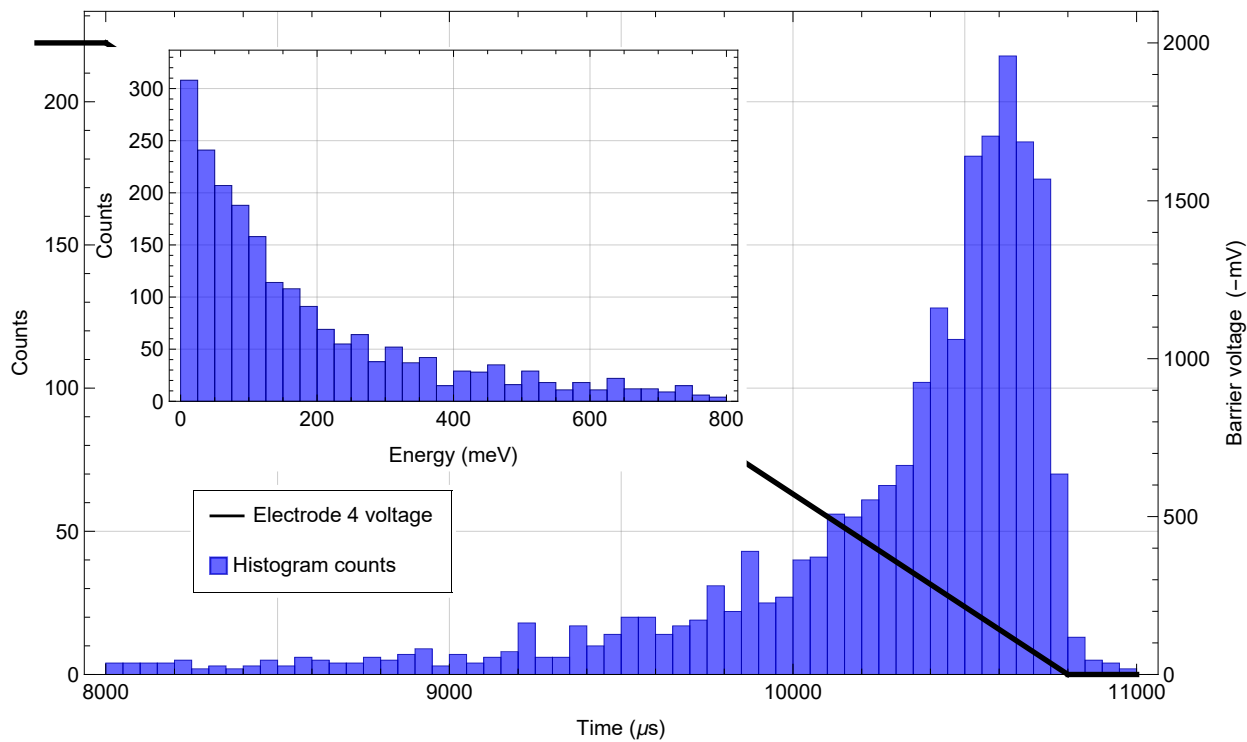

Fig. 5. Histogram of electrons and anions released from the trap. The figure shows the histogram (in blue) from final slow linear dump of electrons and anions (about 1/4 e⁻ and 3/4 H⁻) from a trap between E2–E4 after similar manipulations as in Fig. 4. This data represents the accumulation of 21 runs at a reduced laser energy. The matrix was 80 nm thick. The sublimation heat pulse starts at t=0 and a single ablation pulse is given at t=6140 μs and E2 rises to -5 V from 6200 – 6250 μs closing the trap between E2–E5. The potential well is later compressed between E2–E4 and E4 is brought to -2 V. The inset shows the corresponding energy distribution, with a nearly exponential decay shape, with the counts after the ramp down being incorporated in the first bin (0 – 25 meV).

To determine which species were trapped, we employed a ToF discrimination in the small space available, far from the ideal Wiley–McLaren[42] condition. By trapping the particles in a single electrode, e.g., E2–E4 with E2 at ±5 V and E4 at ±1 V, we can lower E4 to ±0.2 V while raising E3 to ±0.4 V (where the ± applies to positive or negative particles) imposing an acceleration region from E2 to E4 (see Methods for details). The particles are accelerated towards the channeltron and have some short length of free flight. The condition is not ideal but enough to discriminate e⁻, H⁻ and Li⁻/LiH⁻, though not Li⁻ from LiH⁻. Due to the short length of the trap magnet we cannot employ this technique at electrodes farther from the detector. The result of this ToF discriminator is shown in Fig. 6, using the same matrix and ablation conditions as in Fig. 5. The ToF discrimination shows electrons and hydrogen anions as majority, followed by a small fraction of Li⁻ or LiH⁻ for those conditions. For this procedure we had to considerably decrease the ablation laser energy to avoid saturating the detector with the e⁻ signal. Details on the simulation are presented in Methods but it is relevant to mention that it is a simple 1D simulation along the axis; that it does not consider the large finite time for the electrodes switching, which typically has a 1.2 µs time decay; and that it uses an energy distribution that would reproduce Fig. 5.

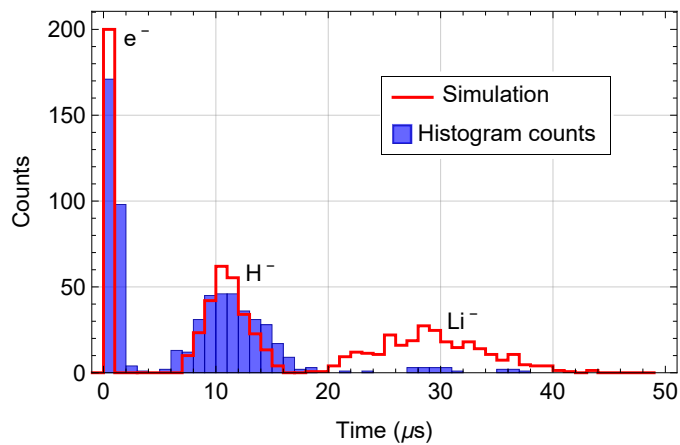

Fig. 6. Time-of-flight discrimination of trapped negative species. ToF discrimination for negative species showing e⁻, H⁻ and Li⁻ (or LiH⁻) in the same trapping conditions and manipulations as in Fig. 5 but now doing a quick ramp down from the trap with E1= -10 V, E2 = -5 V, E3=0 V, E4= -2 V to an acceleration ramp with E1=-10 V, E2= -5 V, E3 = -0.2 V, E4= -0.1V and E5 = 0 V. The experimental data from an accumulation of 21 runs is shown in blue while a simulated histogram for the species identified is shown with a red envelope. In this figure, t=0 is the beginning of the dump, taken as instantaneous in the simulation but lasting over 1.2 µs. The ablation laser energy was considerably decreased to avoid saturation of the CEM with the e⁻ signal. See section Results for more discussion and Methods for the simulation.

Due to some saturation of the CEM with the electron signal in the ToF, we employed another method to get a ratio of H⁻ to e⁻ under these conditions. The electrons arrive at the trap region almost immediately after the ablation in this variation while the H⁻'s take about 100 µs. By measuring that the electrons trapped number would not change from 60 to 90 µs, after ablation and before the arrival time of the H⁻, we can compare the trapped distribution for slow ramps for samples comprised only of electrons to samples trapped at the best time for H⁻. The result is shown in Fig. 7. The ratio of the areas (~3.6) represents that, at its optimal timing, H⁻ is trapped 2.6 times more efficiently than e⁻ and with a similar energy distribution.

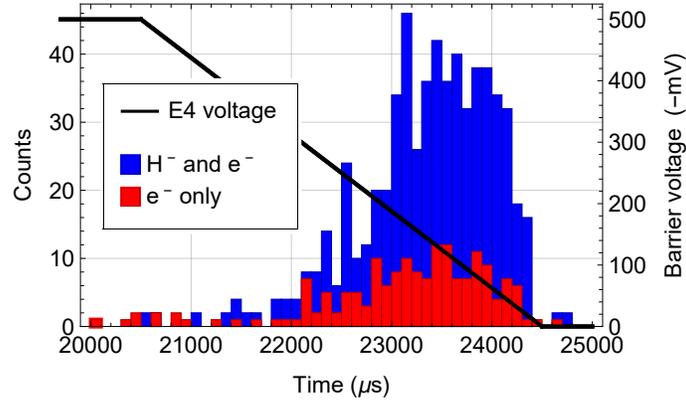

Fig. 7. Trapped electrons and negative hydrogen plus electrons in slow dump regime. Comparison of slow dumps for a sample trapped at earlier times (60–90 µs after ablation) comprised only of e⁻ (in red), to sample comprised of e⁻ and H⁻ (in blue) trapped at the proper H⁻ trapping time 100 µs after ablation. In these conditions, the ratio of integrated counts results in 2.6 times more H⁻ than e⁻ trapped at the optimized H⁻ trapping time. The energy distributions are similar.

Another interesting manipulation, able to discriminate masses, is to make the trap unstable to certain species by changing the electrostatic potentials and the magnetic field. For a harmonic Penning trap to be stable[43] the cyclotron frequency, $\omega_c = q\frac{B}{m}$, and the axial frequency, $\omega_z \sim 1/\sqrt{m}$, should relate as $\omega_c > \sqrt{2}\,\omega_z$. The condition relates the confining potential shape (in $\omega_z$), the magnetic field ($B$) and the particle mass ($m$) and charge ($q$). Compressing and deepening the potential well – changing to -10 V, +10 V and -10 V in electrodes E3, E4 and E5, respectively – we can easily make the trap unstable for species other than e⁻ and H⁻ even at the maximum current of 100 A in the trap magnet. Under this quasi-harmonic trap configuration even H⁻ becomes unstable under ∼ 55 A and we performed a study on the trapped particles number as function of the current. Due to a hardware limitation, we could only set the trap magnet current just before sublimation and could not separate the guiding and initial trap loading processes from the trap instabilities. To keep the magnetic field guiding lines configuration unchanged, the sapphire magnet current was changed in the same proportion as the trap magnet current. In Fig. 8 the result of this study is shown.

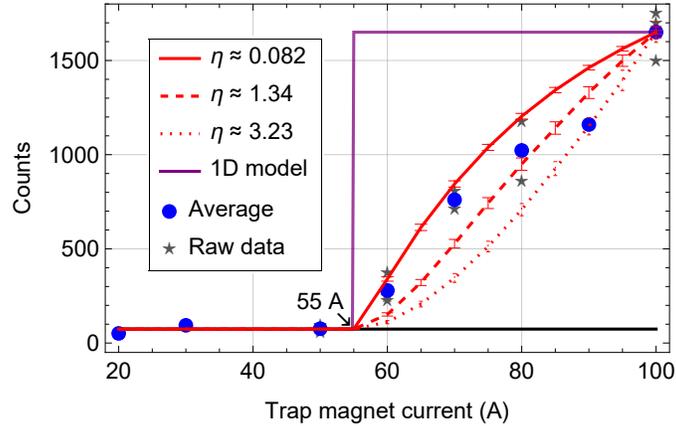

Fig. 8. Trap loading and survival of H⁻ against current in the trap magnet. The data is shown as grey stars, representing individual realizations, and the blue dots are the average of the grey stars values for the corresponding current. Two models are shown. The purple line represents a simple axial 1D model concerning only the (harmonic) axial and cyclotron frequencies relation for stability. The red lines guide the eyes through the results of a Monte-Carlo model considering the entrance aperture of trap region, at magnetic fields much lower (~1/13) than in the trap, and the survival of the particles in the trap given its diameter for different radial energy distributions. The axial energy distribution is taken as an exponential decay as in Fig. 5 and the transverse energy distributions are taken as exponential decays but with different energy scales. In the continuous, dashed, and dotted red lines the ratio of the transverse to the axial energy scales input trial to the simulation is $\eta$ = 1/12, 2, and 20, respectively, while the trapped sample ends with the $\eta$'s shown in the figure. Magnetic mirroring effects are taken into effect in the energy rescaling between the trap entrance aperture to the trap center. The models do not include a full particle tracing but only consider the particles at the entrance aperture and the trap center. The error bars from the simulation represent one standard deviation for the 50 realizations. See section Results and Methods for further discussion.

The data clearly shows a decrease of the number trapped as the current decreases until a cutoff near 55 A where H⁻ becomes unstable and the residual counts at lower currents are due to e⁻. Two models attempt to capture the qualitative behavior. A first one, in purple, supposes a simple axial 1D model for ions already in a harmonic trap resulting in a step function separating the regions of stability and instability. The other considers the entrance aperture transmission of the trap, and trap stability also considering the finite electrodes diameters. At the entrance aperture the magnetic field is small, at a ratio of ~1/13 of that in the trap center, and the radius of curvature of the cyclotron motion may exceed the entrance opening. Inside the trap one may have instability or simply a large cyclotron plus magnetron motion that reaches the electrode walls and lead to losses. Monte-Carlo simulations, using an axial energy distribution like the exponential one in Fig. 5 inset (despite different conditions and manipulations for this data set to that of Fig. 5) and different perpendicular energy distribution lead to the other curves in Fig. 8. The details on the models are presented in Methods.

The good agreement of the second model, without adjustable parameters other than the arbitrary scaling number for the total counts at 100 A, suggests that this method may be further developed as a diagnostic tool for the radial density distribution – requiring validating studies with an imaging detector, or employing an axially displaced photodetachment laser beam, and higher data statistics – or the transverse energy distribution. These are two identified parameters affecting the curves.

The results shown in Figs. 2–8 above represent a subset of the possibilities with this technique. For instance, we produced trapped H⁻ from ablation on $TiH_2$ and performed a few runs with a $H_2$, instead of Ne, matrix. In the present trap we can make all atomic masses above 7 (Li), or higher, to be unstable. Light particles, like electrons can be expelled from the trap by a fast (~1 μs, in the case of e⁻) opening–and–closing of the trap barrier, but this required adding some extra complexity to our hardware and software and we performed just a few validating tests. Therefore, one could select a range of masses to keep in the trap for studies such as molecular formation. Splitting the ablation laser beam, one could simultaneously ablate different targets to produce molecules, or one could press a sintered composite material with the different precursors into a single ablation target. We have also produced a cold beam of electrons from electrostatic sublimation, i.e., we can switch on and off a beam of electrons, and subsequently trap them, just by applying a potential to the sapphire substrate without sublimating a matrix that was previously implanted via laser ablation.

**Discussion**

In a next generation of the apparatus, with a longer trap magnet – and a more uniform field – and a larger number of electrodes we will be able to achieve better mass discrimination and to stack samples in a simple manner: trap in electrodes to the left and move and stack in the electrodes to the right and repeat. Using the described MISu variation we were able to trap over one thousand H⁻ ions after a single ablation pulse. With a diode pumped ablation laser operating at high frequency during a single matrix sublimation and the use of a stronger magnet the system yield can be scaled up by orders of magnitude in number of trapped ions. Trapping heavier species at large numbers will also require higher magnetic fields for guiding them from the sapphire to the trap. We plan to implement a simultaneous detection of positive and negative particles and optical access for photodetachment and $Ca^+$ spectroscopy lasers.

The trap holding time was limited by the heating of the trap magnet at ~1 s time scale. In a test, we held a sample for up to 1.5 s obtaining ~1000 counts in the final dump of that realization. In a future setup we plan to use a superconducting magnet which, together with a better thermal anchor in the trapping region, should enable long holding times and thus to achieve thermalization and perform evaporative cooling of the trapped samples reaching a few kelvins of temperature.

The system has many adjustable parameters, such as matrix growth, ablation, sublimation, and timing. With appropriate parameters we were able to achieve a reliable operation. There are still many possibilities to explore systematically, such as the application of retarding (or accelerating) potentials to the sapphire during sublimation and the use of a $H_2$ matrix. We could only perform very preliminary – encouraging, but not conclusive – explorations on those possibilities.

Moreover, the system still holds many intriguing and open questions. Why we trap only few $H^-$ ions in the typical MISu procedure? Is the $H^-$ not penetrating the matrix, or being stripped of its extra electron as it enters the solid Ne, or being ionized under the matrix's intense electric fields?
We want to explore the conditions to produce ions at the typical temperatures observed for neutral atoms produced by MISu, in the few-kelvin range. The MISu technique generates an intense Li beam at a few kelvins[40], and below, despite Li having a positive (expelling) energy[44] of 0.25 eV (~2900 K) in Ne. This is a strong argument why we expect to produce ions coming from the matrix at such low temperatures as well, even with space charge effects with the ions. The trap closing procedure itself imparts large energy to a fraction of the ions under the rising electrodes potentials and this is clearly seen in Fig. 3. If the trapped ions would exchange some energy, the higher than expected energy of our samples could be coming from the dynamic trapping procedure (one gets 1.6 eV average energy from a trap switching to 5 eV, supposing a uniformly distributed flux of ions over the electrodes region). In the contrary direction, the ratio of magnetic fields of 41 mT below the sapphire, where collisions with the Ne gas may not be relevant anymore, to 92 mT in the trap gives some artificial cooling of the measured axial energy since the transverse energy for particles following the field lines will scale with the magnetic field at the expense of the axial energy. We plan to use laser spectroscopy on suitable cations, such as $Ca^+$ to study the real temperature of the ions' sample and verify at which step of the procedure the sample is gaining energy.

In the present setup there is some Ne reaching the trap region. This condition can be mitigated. Since the system generates positive and negative particles, we could setup the system with different ablation targets and a trap configuration for positive and negative species trapped side–by–side ready to study molecular recombination, as it is done with $\bar{\text{H}}$ formation by mixing $\bar{\text{p}}$ into a e$^+$ cloud.

With the electrostatic sublimation of electrons from the matrix, we speculate on producing a low energy spin–polarized beam of electrons. If the matrix is at 3 K and would be subjected to a 4 T field, a spin thermalized electron sample could be 80% polarized. Such a simple low–energy polarized electrons' source could be very attractive to biochemical experiments, especially with respect to the issue of biology homochirality[45,46], and to fundamental quantum collisional processes.[47]

We believe the data is quite a strong proof–of–principle for what is needed to load H atoms in the $\bar{\text{H}}$ ALPHA experiment trap. The system here presented is at a more advanced stage than other alternatives[48,49] proposed. As for applications with large amounts of ions, such as with tritium or deuterium, we believe the system is scalable. The use of higher power ablation lasers operating at high rates, together with stronger guiding magnetic fields, would improve the numbers by orders of magnitude.

**Conclusions**

We have demonstrated a platform to produce cold beams or trapped samples of electrons, anions, and cations. The system was first designed to produce H$^-$ towards loading the $\bar{\text{H}}$ trap in the ALPHA experiment at CERN and we find it meets the necessary criteria in terms of numbers, energy range, and UHV conditions after trapping. The system should work equally well for T$^-$/T, perhaps suitable for a neutrino mass experiment, and for D$^-$/D for studies relevant to fusion research. Trapped anions and cations, or in a beam, can be used to study chemistry at astrophysical conditions, such as in star formation. The system can generate an electrically controlled electron beam which we speculate might be suitable for spin polarization with applications towards studies of chirality with relevance to biochemistry. Based on the present data and previous studies on MISu, the system is quite versatile and generic in the sense of being able to produce different species of atomic and molecular neutrals, anions, and cations.

**Acknowledgements** This work was supported by: CNPq, FAPERJ, and RENAFAE (Brazilian agencies). We thank Profs. Massimo Xi Liu, from USP-São Carlos, and Vitoria Barthem, from UFRJ, for depositing the NiCr resistive films, first and second batches respectively. We acknowledge participation of Emily Costa in some early experimental runs.

**Author Contributions** The initial conception was proposed by CLC. All the authors participated in the design and implementation at various stages. LOAA, primarily, and RJSC were responsible for the experimental runs.

## Methods

### Matrix Isolation Sublimation (MISu) method

In MISu, a solid matrix is formed by condensation of the inert gas — Ne or $H_2$ — onto a cold sapphire substrate. Atoms and molecules are implanted in the matrix via laser ablation of a solid precursor containing the atomic species of interest. The inert gas matrix together with the atomic and molecular species is sublimated into vacuum by a heat pulse applied on the sapphire. The properties of the atomic beam can be tuned mainly by the thickness of the matrix, the number and energy of ablation pulses, and the power applied to sublimate the matrix[40]. The beams of atoms and molecules can be produced in a fast pulse (in ~10–100 μs) or up to a quasi–continuous regime (in ~1–10 s), depending on the sublimation parameters and matrix size. Pellets of LiH, Ca, Cr, graphite, and other, have been used as ablation targets. As previously reported[41], we produced $Li_nCa_m$ (and $Li_nC_m$) molecules from alternating laser ablation of a Li and a Ca (C) pellets demonstrating the technique's versatility.

**Experimental Setup and procedure**

The sapphire substrate, upon which we deposit the neon matrix, is coated on both sides with NiCr resistive films. The back film is used as a resistor to apply a sublimating heat pulse while the front film serves as a mirror and provides a way to bias the sapphire with a positive or negative voltage. The bias can be used to repel or attract charged species to the matrix during the ablation process and/or during the sublimation phase, but all the data presented here had no bias with the front NiCr film grounded.

In Fig. 1a we present the basic setup used in this study. A two–stage closed cycle cryostat with 1 W of cooling power at 4 K is used. A UHV environment can be achieved below 6.9 K, where the saturated vapor pressure of Ne is predicted to be ~ $10^{-7}$ Pa, reaching ~$10^{-19}$ Pa at 4 K. This UHV condition is not met during the matrix growth or sublimation, and it takes milliseconds for the Ne gas to be cryopumped back to the walls. Due to a poor thermal link in the present setup, the trap electrodes region only reaches ~6.5 K while the main chamber is at ~4 K. The sapphire substrate is thermally anchored to the 4 K stage through a designed thermal link that allows it to reach 3.2 K and be heated for the sublimation process without warming the whole experimental cell. The Ne gas delivery tube is brought into the cryostat with a first thermal link at the 1st stage (40 K nominal) and its exit is above 16 K, to avoid plugging, and directed towards the sapphire. The laser reflections on the interfaces, vacuum–matrix and matrix–sapphire, cause interference fringes that allow monitoring the matrix thickness down to the atomic monolayer as well as performing Doppler–sensitive spectroscopy of neutral Li atoms as they sublimate from the matrix into vacuum.

During or after the matrix growth a single, or multiple, laser ablation pulses onto the appropriate solid precursor – mainly LiH during this study – release and implant neutral and charged particles into the matrix or reflect from it. The matrix can be sublimated at different times with respect to the ablation pulses. We have employed many variations on these timing parameters and laser pulse energies. For ablation we mostly employed a doubled Nd:YAG at 532 nm, with a few mJ in 5 ns pulses. This flashlamp pumped laser became unreliable and we switched to a diode laser pumped Nd:YAG at 1064 nm with up to 1.5 mJ using a tighter focusing for similar fluences. The front sapphire terminal can be monitored for the charge deposited into (or released from) the matrix, acting as a capacitor, during the ablation (or sublimation) process. Through laser ablation, enough charge can be deposited to achieve many volts of space charge or to thermally explode the matrix. Both the ablation process and space charge variations can affect the results. We monitored the matrix size, the deposited charge, and the sapphire temperature change due to the ablation and sublimation. Adjusting these parameters and the very critical timing for switching the trapping electrodes potentials, we have conducted many experimental cycles exploring different parameters and establishing reproducibility under the same conditions. The goals were to maximize the number of trapped ions at low ablation energies while minimizing their energy after capture.

A superconducting magnet (see Fig. 1a) placed around the sapphire generates a field of ~41 mT in a configuration to guide the particles into the trap region. The Penning–Malmberg trap, in the horizontal direction, is composed of six similar electrodes, named E0–E5, of 17.4 mm of inner diameter and 20 mm length each. The magnetic field at the trap is generated by an external resistive magnet which produces 92 mT on axis, around E3–E4, at 100 A. Such a short magnet was dictated by space constraints and resulted in a highly non-uniform field. This leads to magnetic mirroring reflections if we dump the particles from electrodes E2–E3. Due to heating, the magnet can only be left energized for ~1.5 s, limiting the trapping time. The resulting field from the two magnets generates a guiding magnetic field that leads the particles from the source making a turn into the trap. Along the guiding lines, the field reaches values as low as 1.5 mT, enough for guiding low energy and low mass species.

For detection of the charged species, we use a Channel Electron Multiplier (CEM), a Magnum Channeltron from Photonis Inc, to the right of electrode E5 in Fig. 1a,b. We set the voltages in the CEM to attract either negative or positive charges before each experiment. For cations, the cone of the CEM is set at -2 kV while the anode is grounded via a load resistor of 1 kΩ. For anions, the cone is set at +0.4 kV while the anode is connected to +2.4 kV via a 1.2 MΩ resistor. In both configurations the output is capacitively coupled out of the cryostat and into a radio–frequency pre–amplifier which is directly recorded by an oscilloscope at 40 GSa/s, with up to 5 ms acquisition, and analysed by a peak detection routine above a certain threshold. To prevent excess black–body heat coming from the CEM into the trap we partially anchor the wires to the CEM at the 40 K heat shield, at the known expense of a higher detector's resistance resulting in longer recharging times. With a high detection rate, we had to deal with saturation of the cold detector. Particularly, using a fast (~1.2 µs) dump configuration for a time–of–flight mass discrimination (ToF) for negative particles, the first coming electron signal can saturate the CEM that ends up detecting very few H⁻ afterwards.

**Time-of-Flight mass discrimination (ToF-MD) of trapped species**

Once the ions are trapped, we have implemented a potential ramp to discriminate the masses. A Wiley-McLaren[42] configuration would require a free-flight region twice as long as the accelerating region. Within our spatially constrained trap, we could switch from a trap configuration as shown in Fig. 9 below.

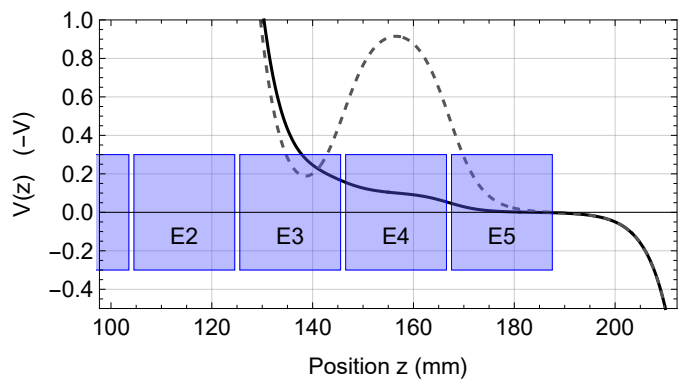

Fig. 9. Time-of-flight mass discrimination potentials for anions. The dashed curve is the initial trap around E3 before the potentials are switched to the ToF configuration shown in solid line. The negative potential at z > 200 mm is due to the CEM.

For initial estimates of this ToF-MD functionality we used a simple analytical function that approximately simulate each electrode potential and added their values to represent different trap configurations. The simplest function, with just 3 adjustable parameters, is given by:

$$V(z) = \sum_{n=0}^{5} f_A A_n \left\{ \text{Erf}\left(\frac{dz+(z-z_{0n})}{\delta z}\right) + \text{Erf}\left(\frac{dz-(z-z_{0n})}{\delta z}\right) \right\},$$

where $A_n, z_{0n}$ represent the applied voltage amplitude and the center position of the $n^{th}$ electrode. The geometrical factor $f_A \approx 0.47$, $\delta z \approx 7$ mm, and $dz \approx 11$ mm is the length of the electrode (10 mm) plus the distance between electrodes (1 mm). From the above expression the electric field can be analytically obtained.

The configuration for the ToF-MD, with precise potentials, is shown in Fig. 9 below. Notice that the electric field is not ideally uniform in the acceleration region, and the free-flight region is small compared to the acceleration one.

The simulation considered a 1D system, along the trap axes, concerning only the electrostatic field produced by the electrodes and CEM detector and not the magnetic field. According to the energy distribution in Fig. 6, we place ions at the returning points in the trap, in both ends, given its energy and let each ion evolve inside the trap for a random time between 0 and half of the period of axial oscillation for that energy. This new state of each ion is the corresponding initial state for the second part of the simulation where the trap is switched instantaneously to the ToF configuration. The actual driving hardware takes about 1.2 µs to reach 90% of the set value and this would cause extra spreading in the detection time distribution, but it was not considered in the simulation. The results are sufficient for discriminating our species of interest for this work.

For the ToF-MD simulations shown in Fig. 6 we used an interpolation for the electrodes' voltage configuration, $V_{ToF-MD}(z)$, given by a finite element package solver instead of the analytical potential described above. The time-of-flight is given by the usual integration $t_{ToF-MD} = \int_{z_0}^{z_f} \frac{1}{\sqrt{\frac{2}{m}\{\epsilon - q\,V_{ToF-MD}(z)\}}} dz$, where $z_0$ is the position at the time of switch, $z_f$ is a position near the CEM – where particles at rest would take less than 0.2 µs to reach the CEM – and $\epsilon$ is the total energy the particle has immediately after the switch, i.e., its initial kinetic energy (when it was in the trap) plus the potential energy in ToF configuration: $q\,V_{ToF-MD}(z_0)$.

**Magnetic field variation study**

To analyze the stability of H⁻ and the losses by collisions in the trap entrance as we change the trapping/guiding magnetic field, we repeated for different magnetic field conditions the following steps: a matrix of about ~370 nm is grown; the currents on the magnets are set to the desired value at maximum, 100 A (2.5 A) in the trap (sapphire) magnet; during a slow sublimation, a single laser ablation hits the LiH target; the trap loading potential, with the electrode configuration (in V) E0=0, E1=0, E2=-5, E3=0, E4=0, E5=-5, is closed 100 µs after the ablation; the sample is adiabatically compressed by changing the electrodes configuration and waiting for 1 ms; particles with energies above ~700 meV are released; after 500 µs the remaining trapped particles are slowly released and detected. The data is shown in Fig. 8.

Some assumptions were made for constructing the models (1D and 3D) to qualitatively describe the survival of H⁻ after this process. A uniform magnetic field along the entire trapping volume was considered, as the trap is just one electrode wide, taking the mean value of field in the region. We considered the trap to be harmonic for energies < 700 meV since the axial oscillation frequency changes less than 2% within this energy range, hence the equations for the stability inside the trap are well-defined. The simple axial 1D model only considered the stability criteria of the relation of the cyclotron and axial frequencies. For the 3D models, we employed a Monte-Carlo simulation considering an axial (z) exponential decaying energy distribution as measured in Fig. 5. The transverse energy distribution was also considered exponential but with different decay values and it was initially mapped only on $v_y$. The initial position of the anions in the $xy$-plane inside the trap $(x_0)$ imitates a uniforme surface distribution but mapped only in $x$ with probability proportional to $x$.

The axial and transverse energies are drawn for a particle inside the trap. Considering the transverse energy adiabatic propagation in magnetic field – where the transverse energy scales with magnetic field at the expense of axial energy – we map both the transverse and axial energies to the entrance ($\epsilon_{t,\text{in}}$ and $\epsilon_{z,\text{in}}$) of the trap region (just left of E0). If $\frac{\epsilon_{z,\text{in}}}{\epsilon_{t,\text{in}}} > \frac{B_{\text{trap}}}{B_{\text{in}}} \approx 12.85$, where $B_{\text{trap}}$ and $B_{\text{in}}$ are the magnetic fields at the trap and at the entrance, the particle is considered in the simulation since it will not reflect magnetically. If $\epsilon_{t,\text{in}} > 172$ meV the cyclotron radius is greater then the entrance radius and we consider the particle lost at the entrance. If the particle enters the trap, given the initially drawn position $(x_0)$ and transverse energy inside the trap ($\epsilon_{t,\text{trap}}$, supposed initially along $y$), we calculate the maximum radius[50] of the particle as the sum of the magnetron and cyclotron radii for this particle for each current considered. If this maximum radius, for each current, is larger than the electrode radius this particle is lost in the trap. Finally, if the current is bellow ~55 $A$, the trap is unstable and no H¯ would be trapped.

The shapes of the curves resulting from these simulations have a dependence with the transverse energy distribution (see Fig. 8). Fig. 10(a,I-c,III) shows the histograms of distributions for the particles that survived the losses considered for the trap at 100 A consisting of the transverse and axial energies ($\epsilon_{t,\text{trap}}$ and $\epsilon_{z,\text{trap}}$) distributions and initial position ($x_0$) considering three different initial distributions for the transverse energy. Notice that a much lower value of transverse energy remained in the trap compared to large trial value considered in Fig. 10(c,I).

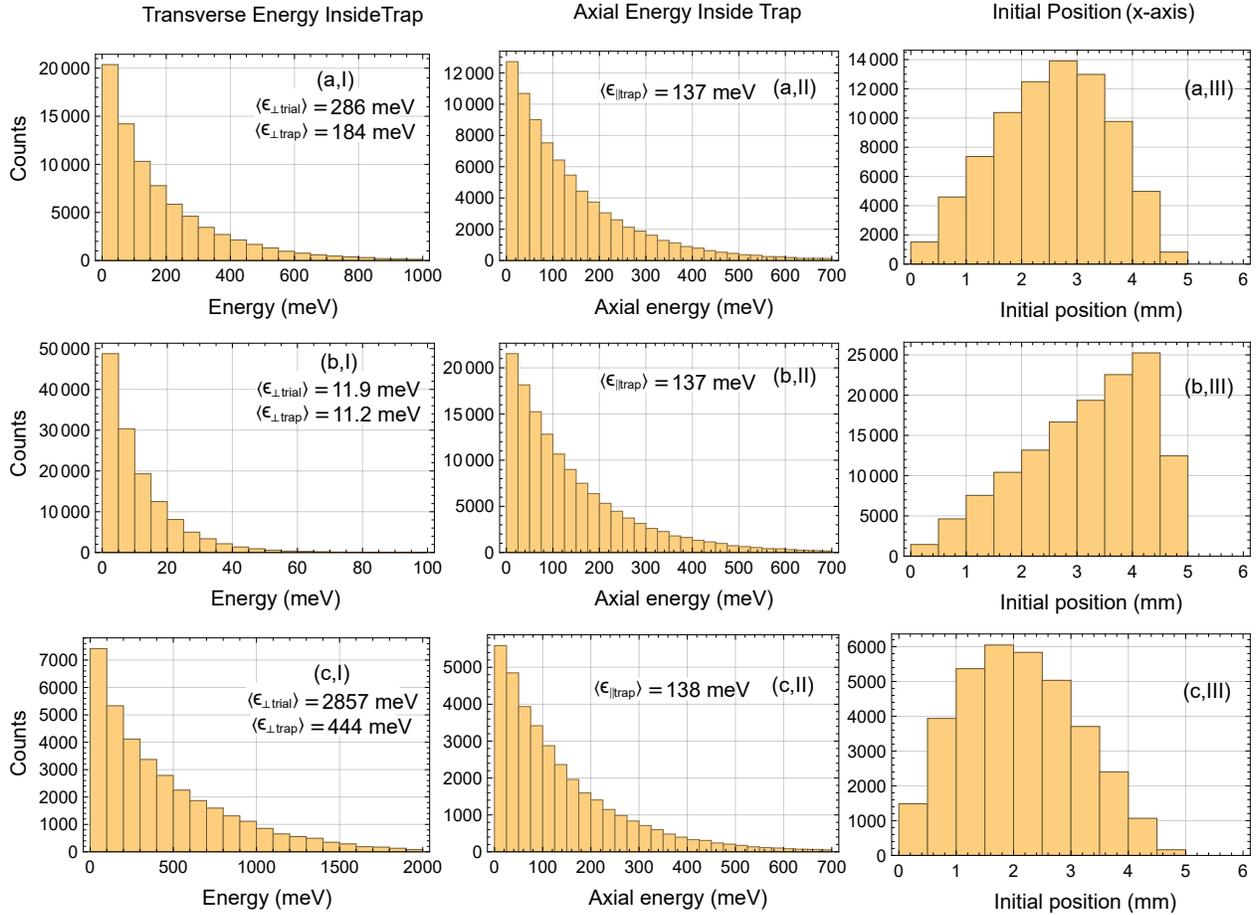

Fig. 10. Monte-Carlo simulations for the surviving species in the trap. Histograms from the Monte-Carlo simulations on the surviving particles inside the trap for a trap magnet current of 100 A for the 3D model with different input trial exponential energy scales for the transverse energy distribution: (a) 2 times greater than in the axial distribution, (b) 12 times smaller and (c) 20 times greater. The average energy values for trapped particles are shown in (a,I) – (c,II). In all three cases an exponential decaying axial energy distribution was assumed in the beginning, but some low energy particles are magnetically reflected and thus considered not trapped. Notice the very large loss at the highest transverse energy distribution from the total of 150000 particles, simulated at each case.

**Data availability**

The datasets generated during and/or analysed during the current study are available from CLC or LOAA (claudio.cesar@cern.ch, ldeazevedo@if.ufrj.br ) on reasonable request.